\documentclass[reprint, amsmath, amssymb, aps, superscriptaddress, prx]{revtex4-2}
\usepackage{graphicx}
\usepackage{multirow}
\usepackage{dcolumn}
\usepackage{bm}
\usepackage{hyperref}
\usepackage[mathlines]{lineno}
\usepackage{physics}
\usepackage{float}
\usepackage{diagbox}
\usepackage{booktabs}
\usepackage[normalem]{ulem}
\usepackage{resizegather}
\usepackage{newtxmath}

\hypersetup{
    colorlinks=true,
    linkcolor=red,
    filecolor=magenta,
    citecolor=blue,
    urlcolor=blue,
}

\newcommand{\hide}[1]{}

\newcommand{\IITB}{Department of Physics, Indian Institute of Technology Bombay, Powai, Mumbai, Maharashtra 400076, India}
\newcommand{\Quicst}{Centre of Excellence in Quantum Information, Computation, Science and Technology,
Indian Institute of Technology Bombay, Powai, Mumbai, Maharashtra 400076, India}
\newcommand{\SNBose}{S. N. Bose National Centre for Basic Sciences, Block JD, Sector III, Salt Lake, Kolkata 700 106, India}
\newcommand{\ETH}{Institut f\"ur Theoretische Physik, ETH Z\"urich, Wolfgang-Pauli-Str. 27 Z\"urich, Switzerland.}

\begin{document}

\title{Thermodynamic Probes of Multipartite Entanglement in Strongly Interacting Quantum Systems}

\author{Harsh Sharma} \affiliation{\IITB}

\author{Sampriti Saha} \affiliation{\IITB} \affiliation{\ETH}

\author{A. S. Majumdar}\affiliation{\SNBose}

\author{Manik Banik}\affiliation{\SNBose}

\author{Himadri Shekhar Dhar} \affiliation{\IITB}\affiliation{\Quicst}


\begin{abstract}
Quantifying multipartite entanglement in quantum many-body systems and hybrid quantum computing architectures is a fundamental yet challenging task. In recent years, thermodynamic quantities such as the maximum extractable work from an isolated system (the ergotropy) have allowed for entanglement measures that are operationally more accessible. However, these measures can be restrictive when applied to systems governed by Hamiltonians with strong collective or interparticle interactions. Motivated by advances in quantum simulators, we propose a framework that circumvents these restrictions by evaluating global and local ergotropy either through controlled quenching of interactions or by measuring suitable local observables only. We show that this formalism allows us to correctly estimate genuine multipartite entanglement in both stationary and time-evolved states of systems with strong interactions, including parametrized quantum states simulated on a quantum circuit with varying circuit depth and noise. We demonstrate its applicability to realistic physical models—namely, the Tavis–Cummings model, the three-level Dicke model, and the transverse-field Ising model—highlighting its potential as a versatile tool for characterizing entanglement in near-term quantum simulators.
\end{abstract}

\maketitle

\section{Introduction}
Entanglement lies at the heart of quantum theory, distinguishing it fundamentally from classical physics~\cite{Horodecki2009,Erhard2020,Terhal2025}. Its conceptual origins can be traced to the seminal works of Einstein–Podolsky–Rosen and Schrödinger in the 1930s~\cite{Einstein1935,Schrodinger1935}, and was later placed on firm experimental footing through Bell’s theorem~\cite{Bell1964} and its landmark tests~\cite{Freedman1972,Aspect1982,Weih1998}. The recognition of entanglement as a tangible physical resource, however, emerged much later—most notably through its pivotal role in quantum cryptography~\cite{Ekert1991} and quantum communication~\cite{Bennet1992,Bennet1993,Buhrman2010}. Since then, entanglement has become central to the development of quantum information processing and computation~\cite{Raussendorf2001,Knill2001,Briegel2009,Nielsen2010}, while also providing deep insights into diverse areas of modern physics, from quantum many-body systems~\cite{Amico2008} to black hole thermodynamics~\cite{Das2010,Solodukhin2011,Belfiglio2025}. Moreover, it underlies a range of collective quantum phenomena, including superradiance~\cite{Lambert2004}, quantum phase transitions~\cite{Osborne2002,Osterloh2002,Vidal2003}, and field-theoretic correlations in quantum critical systems~\cite{Calabrese2004}.

Despite its utility, reliably quantifying entanglement remains a formidable challenge. Even for bipartite systems, entanglement measures for general mixed states in higher dimensions typically lack closed form expressions and require nontrivial numerical optimization~\cite{Karol1999, Audenaert2001, Ryu2008, Beat2009, Allende2015, Toth2015, Arceci2022}. The difficulty escalates significantly in multipartite systems, where the classification of entanglement becomes more complex due to the existence of inequivalent entanglement classes~\cite{Bennett2000,Osterloh2006,Guhne2009,Walter2016}. Of particular interest is genuine multipartite entanglement (GME), which characterizes states that are inseparable across all possible bipartitions~\cite{Horodecki2009,Walter2016}. Several approaches -- including entanglement witnesses~\cite{Guhne2002}, entropy based criteria~\cite{Huber2010,Ma2011,Beckey2021}, and geometric approaches~\cite{Shimony1995,Barnum2001,Wei2003,SenDe2010} -- have been developed to detect and quantify GME. However, these techniques are often computationally demanding or tailored to specific families of states, limiting their applicability in complex interacting systems.

An intriguing and operationally motivated alternative arises from quantum thermodynamics~\cite{Campaioli2018}. Building on the early connections between entanglement and thermodynamics~\cite{Horodecki2002}, thermodynamic quantities such as ergotropy—the maximum work extractable from a quantum system under entropy-preserving operations~\cite{Allahverdyan2004,Alicki2013}—have found relevance in entanglement quantification. In particular, the ergotropic gap, defined as the difference between the extractable work under global and local unitary operations, provides a way to quantify quantum entanglement and other nonclassical correlations in bipartite states~\cite{Mukherjee2016,Alimuddin2019,Alimuddin2020,Alimuddin2025}, making it an experimentally viable indicator of entanglement \cite{Joshi2024a,Joshi2024b,Hoang2024}. More recent studies have extended this approach to multipartite systems, offering means to quantify both multipartite entanglement and GME~\cite{Puliyil2022,Yang2023a,Mula2023,Yang2024,Sun2024}, thereby linking the gains in work extraction to underlying quantum correlations. These methods are particularly appealing due to their operational significance and the fact that they can be implemented without full state tomography. Nevertheless, their effectiveness diminishes in systems with strong interactions or couplings, where the ground state is not necessarily separable and local operations fail to capture global correlations accurately (see Ref.~\cite{Puliyil2022}).

\begin{figure*}
\centering
\includegraphics[width=0.9\linewidth]{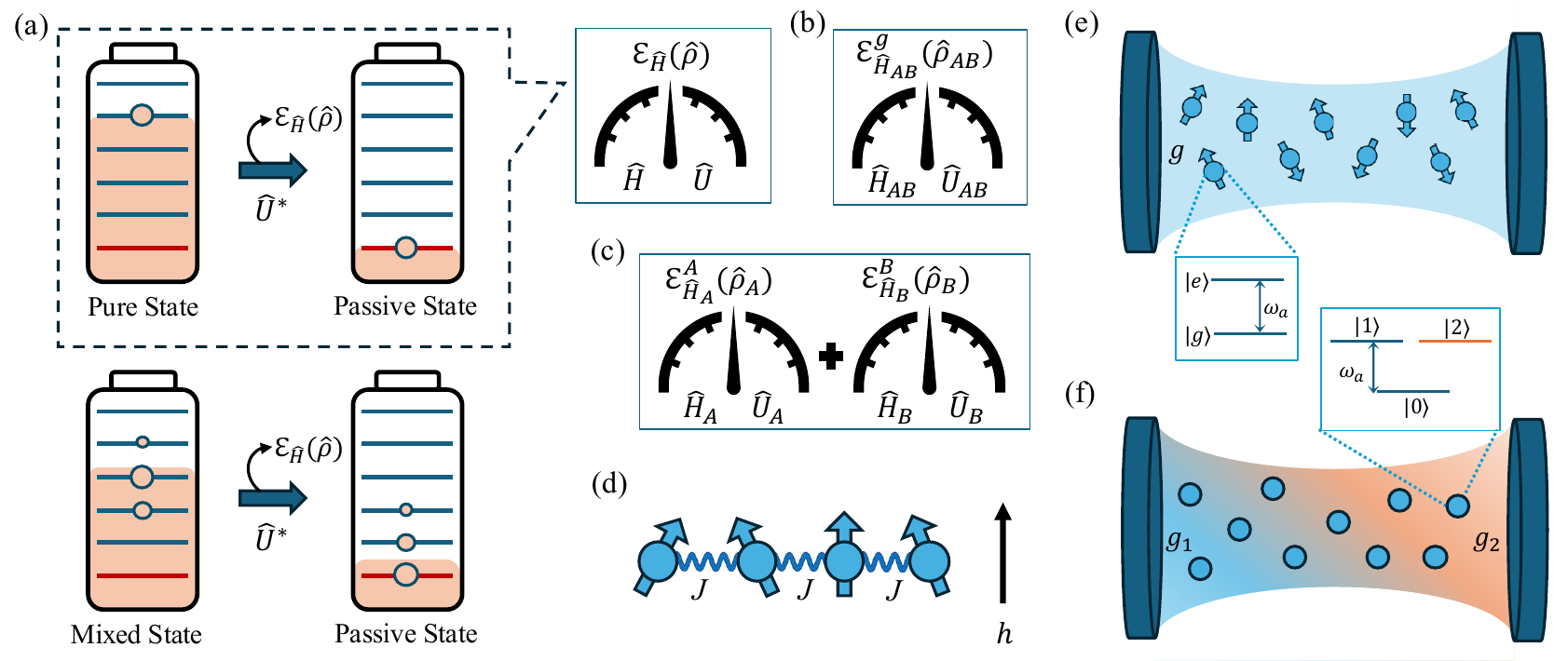}
\caption{Visualization of ergotropy as the maximum work that can be extracted from a charged quantum battery. {(a) The optimal unitary $\hat{U}^*$  discharges the battery to the passive state, which is the lowest energy state with same entropy. This is the ground state if the initial state is pure. However, for mixed initial states, the passive state is not the ground state but a mixture of low energy states. The blue box represents a device that measures the ergotropy $\mathscr{E}_{\hat{H}}(\hat{\rho})$, optimized over a set of unitaries $\hat{U}$ and Hamiltonian $\hat{H}$.
(b) For a state $\rho_{AB}$, the global entropy $\mathscr{E}^g_{\hat{H}_{AB}}({\hat{\rho}_{AB}})$ is measured if the device optimizes over all global unitaries $\hat{U}_{AB}$, whereas
(c) the local ergotropy is measured for the marginals $\rho_{A(B)}$, optimized over all local unitaries $\hat{U}_{A(B)}$ i.e., $\mathscr{E}^l_{\hat{H}_{AB}}({\hat{\rho}_{AB}}) = \mathscr{E}^A_{\hat{H}_{A}}({\hat{\rho}_{A}}) + \mathscr{E}^B_{\hat{H}_{B}}({\hat{\rho}_{B}})$. The local ergotropy expression comes from the simplification of Eq.~\eqref{Eq:locErg} after quenching the interaction.
The figure also illustrates the strongly interacting systems we study: (d) the transverse-field Ising model, (e) the Tavis-Cummings model and (f) the three-level Dicke model.}}
\label{fig:FullErg}
\end{figure*}

In this work, we introduce a versatile framework to quantify ergotropy based genuine multipartite entanglement in systems with strong interparticle or collective interactions.
{Taking motivation from the realm of quantum simulators~\cite{Cirac2012}, where operators and measurements can be controlled, both the
the global and local ergotropy can be captured by quenching the system~\cite{Cheneau2012,Bernien2017} or making local measurements~\cite{Bayat2018,Villa2020}. 
Such an approach allows us to estimate the thermodynamics based properties of the state, independent of the nature of interaction in the collective system, and is broadly applicable, as we demonstrate, across representative models from quantum optics and many-body physics that involve strongly correlated particles.} Specifically, we consider the Tavis–Cummings~\cite{Tavis1968,Kirton2019} and spin-1 Dicke models~\cite{Lin2022,Fan2023}, where two-level atoms or spins are strongly coupled to a photon field. We also study strongly interacting spin systems governed by the transverse-field Ising model~\cite{Gennes1963,Sachdev2011}. These models encompass diverse interaction types and collective excitations relevant in cavity QED~\cite{Walther2006} and spin-lattice systems~\cite{Kogut1979,Latorre2009}. Our framework enables both analytical and numerical characterization of GME in stationary states, revealing correlations that are otherwise difficult to capture. Importantly, our method extends naturally to time-evolved states implemented on quantum circuits. Using a variational quantum algorithm~\cite{Cerezo2021,Bharti2022}, we demonstrate how ergotropy-based GME can be estimated on near-term quantum devices without full state reconstruction. This circuit-level formulation highlights the power, practicality, and experimental relevance of the method, making it particularly suitable for entanglement estimation in current noisy intermediate-scale quantum (NISQ) technologies~\cite{Preskill2018}.

The remainder of the paper is structured as follows. In Sec.~\ref{Sec2}, we review the concept of ergotropy and introduce entanglement quantifiers for strongly interacting quantum systems. Section~\ref{Sec3} presents both analytical and numerical results for evaluating genuine multipartite entanglement across different systems. In Sec.~\ref{Sec4}, we 
describe a variational quantum algorithm to estimate entanglement in time-evolved states of quantum spin models simulated on a quantum circuit. Finally, Sec.~\ref{Sec5} summarizes our findings and discusses future directions.

\section{Ergotopy and multiparty entanglement}\label{Sec2}
Throughout the paper we follow the standard notations familiar in the quantum information community. State of a quantum system is described by a density operator $\hat{\rho}\in\mathcal{D}(\mathscr{H})$ acting on the Hilbert space $\mathscr{H}$ associated with the system; $\mathcal{D}(\cdot)$ denotes the set of density operators. A state is called pure \textit{iff} $\hat{\rho}^2=\hat{\rho}$, and such states can be associated with a ray vectors $\ket{\phi}\in\mathscr{H}$. A state $\hat{\rho}_{N}\in\mathcal{D}(\mathscr{H}_s:=\otimes_{i=1}^N\mathscr{H}_i)$ of a composite quantum system consisting of $N$ subsystems is called fully separable {\it iff} it is a probabilistic mixture of fully product states, i.e. $\hat{\rho}^{fs}\in\mathrm{Conv.Hull}\big\{\ket{\psi}=\otimes_{i=1}^N\ket{\psi_i}~\text{s.t.}~\ket{\psi_i}\in\mathscr{H}_i\big\}$; $\mathscr{H}_i$ denotes the Hilbert space associated with the $i^{th}$ subsystem, and $\mathrm{Conv.Hull}\{\cdot\}$ denotes the convex hull of a set, i.e. the minimal convex set containing the set. Partitioning $N$ subsystems into two non-zero partitions $\mathcal{P}~\&~\mathcal{P}^{\mathrm{c}}$ (i.e. $\mathcal{P}\cap\mathcal{P}^{\mathrm{c}}=\emptyset~\&~\mathcal{P}\cup\mathcal{P}^{\mathrm{c}}=\{1,\cdots,N\}$), a state $\hat{\rho}^{\mathcal{P}|\mathcal{P}^{\mathrm{c}}}$ is called biseparable across $\mathcal{P}$ vs $\mathcal{P}^{\mathrm{c}}$ bipartition {\it iff} $\hat{\rho}^{\mathcal{P}|\mathcal{P}^{\mathrm{c}}}\in\mathrm{Conv.Hull}\big\{\ket{\psi}=\ket{\psi_{\mathcal{P}}}\otimes\ket{\psi_{\mathcal{P}^{\mathrm{c}}}} \text{s.t}~\ket{\psi_{\mathcal{P}}}\in\mathscr{H}_{\mathcal{P}}~\&~\ket{\psi_{\mathcal{P}^{\mathrm{c}}}}\in\mathscr{H}_{\mathcal{P}^{\mathrm{c}}}\big\}:=\mathrm{CH}(\mathcal{P}|\mathcal{P}^{\mathrm{c}})$. On the other hand, a state is called biseparable {\it iff} it lies within the convex hull of the sets $\mathrm{CH}(\mathcal{P}|\mathcal{P}^{\mathrm{c}})$ constructed with all possible non-zero disjoint partitioning of the subsystems. States that are not biseparable are called genuinely entangled. A canonical example of such a state is the $N$-qubit Greenberger–Horne–Zeilinger (GHZ) state $\ket{GHZ}_N:=(\ket{0}^{\otimes N}+\ket{0}^{\otimes N})/\footnotesize{\sqrt{2}}\in(\mathbb{C}^2)^{\otimes N}$ \cite{Greenberger1989}, where $\ket{0},~\ket{1}$ are the qubit computational states. 

Consider a system  of $N$ interacting {quantum particles governed by a Hamiltonian $\hat{H}_{s}$ and prepared in the  state $\hat{\rho}_{N}\in\mathcal{D}(\mathscr{H}_s)$. The ergotropy of the state is defined as the maximum amount of energy (or work) that can be extracted from $\hat{\rho}_{N}$, using only unitary (or entropy preserving) processes \cite{Alicki2013,Campaioli2018}, i.e.
\begin{multline}
\mathscr{E}_{\hat{H}_s}({\hat{\rho}_N}) := \text{Tr}\left [\hat{H}_s \hat{\rho}_N \right]\\ 
- \min_{\hat{U}_s \in \mathscr{H}_s} \left(\text{Tr}\left[ \hat{H}_s \left(\hat{U}_s\hat{\rho}_N\hat{U}_s^\dagger \right) \right]\right),\label{Eq:ErgDef}
\end{multline}
where {the} minimization is over all possible unitary operators $\hat{U}_s$ acting on $\mathscr{H}_s$. Ergotropy thus quantifies the difference in energy between the state $\hat{\rho}_N$ and its passive state $\hat{\rho}^p_N$, which is the transformed, minimum energy state, such that $\text{Tr}[\hat{H}_s\hat{\rho}^p_N]=\min_{\hat{U}_s \in \mathscr{H}_s} (\text{Tr}[ \hat{H}_s (\hat{U}_s\hat{\rho}_N\hat{U}_s^\dagger )])$. 
The notion of passivity becomes more apparent once the Hamiltonian is represented in its spectral form $\hat{H}_s = \sum_{i=1}^d E_i \ket{E_i}\bra{E_i}$, with $E_i \leq E_{i+1}$. Now, arranging eigenvalues $\{\lambda^i_N\}$ of $\hat{\rho}_N$ in decreasing order, the passive state reads as $\hat{\rho}^p_N = \sum_{i=1}^d \lambda^i_N \ket{E_i}\bra{E_i}$ (see Fig.~\ref{fig:FullErg}(a)). In other words, for a passive state,  a lower  energy eigenstate corresponds to a higher  population. Consequently, ergotropy of the state $\hat{\rho}_N$ reads as $\mathscr{E}_{\hat{H}_s}({\hat{\rho}_N}) = \text{Tr}\left[\hat{H}_s(\hat{\rho}_N-\hat{\rho}^p_N)\right]$. Likewise the von Neumann entropy of a quantum state its passive state energy is also uniquely determined for any given state.

For bipartite systems, the notion of ergotropy plays a crucial role in certifying entanglement. For pure bipartite states, the passive-state energy of the marginals equals the ground-state energy whenever the marginal states are pure—that is, when the global state is separable. In contrast, entangled pure states, having mixed marginals, lead to higher passive-state energies, as heuristically illustrated in Fig.~\ref{fig:FullErg}(a). This observation motivates the definition of the ergotropic gap for a bipartite quantum state, quantified as the difference between ergotropies obtained under global and local unitary optimizations~\cite{Mukherjee2016}. 
For a bipartite state $\ket{\psi}_{AB}$, the ergotropic gap is defined as
\begin{align}
\Delta_{\hat{H}_{AB}}^{A|B}(\ket{\psi}_{AB}) &:= \mathscr{E}^g_{\hat{H}_{AB}}({\ket{\psi}_{AB}}) - \mathscr{E}^l_{\hat{H}_{AB}}({\ket{\psi}_{AB}}),
\label{Eq:ErgGap}
\end{align}
where $\mathscr{E}^g_{\hat{H}_{AB}}$ and $\mathscr{E}^l_{\hat{H}_{AB}}$ denote the global and local ergotropies of the state, respectively. The ergotropic gap serves as a faithful quantifier of entanglement~\cite{Alimuddin2019,Alimuddin2020}, satisfying the basic requirements of an entanglement monotone~\cite{Vedral1997}. More recently, for multipartite systems involving more than two subsystems, suitable functions of ergotropic gaps across all bipartitions have been proposed to capture genuine multipartite entanglement (GME)~\cite{Puliyil2022}. However, this formalism remains consistent only for pure states under non-interacting Hamiltonians. In the presence of strong interparticle interactions, the approach breaks down: if the ground state itself is entangled, its mixed marginals cannot be reached by any local unitaries in Eq.~\eqref{Eq:ErgGap}, leading to a non-zero ergotropic gap even for {biseparable states $\ket{\psi}_{AB}$.}

\subsection{{Ergotropic gap for interacting systems}}

The incongruity in computing the ergotropic gap for interacting quantum systems can be resolved through a simple yet powerful extension of the original formalism. The motivation stems from the domain of quantum simulation~\cite{Cirac2012}, where a complex, interacting quantum system is emulated using more controllable platforms such as neutral atoms~\cite{Bloch2008}, trapped ions~\cite{Blatt2012}, or superconducting circuits~\cite{Krinner2019}. In such systems, the energetics of a quantum state {$\ket{\psi}_{AB}$}—whether governed by an interacting Hamiltonian or implemented within a quantum circuit—can be experimentally accessed either by quenching the interactions~\cite{Cheneau2012,Bernien2017} or by measuring only local observables~\cite{Bayat2018,Villa2020}. Importantly, since the ergotropic gap is an intrinsic property of the quantum state, its evaluation remains independent of the specific interaction structure in the Hamiltonian or the measurement scheme.

For instance, an interacting Hamiltonian $\hat{H}_{AB}$ governing the system can be written as 
$
\hat{H}_{AB} = \hat{H}_{A}+\hat{H}_{B}+\hat{H}^{int}_{AB},  
$
where $\hat{H}_A$ and $\hat{H}_{B}$ are the local Hamiltonians corresponding to partitions $A$ and $B$, and $\hat{H}^{int}_{AB}$ denotes the interaction term between them. Using the definition in Eq.~\eqref{Eq:ErgDef}, the global ergotropy thus reads as
\begin{multline}
\mathscr{E}^g_{\hat{H}_{AB}}({\ket{\psi}_{AB}}) := \text{Tr}\left [\hat{H}_{AB} \hat{\rho}_{AB} \right] \\- \min_{\hat{U}_{AB} \in \mathscr{H}_A \otimes \mathscr{H}_B} \left(\text{Tr}\left[ \hat{H}_{AB} \left( \hat{U}_{AB}\hat{\rho}_{AB} \hat{U}_{AB}^\dagger \right) \right]\right),
\label{Erg_global}
\end{multline}
where {$\hat{\rho}_{AB}=|\psi\rangle\langle\psi|_{AB}$} and the second term is the passive state energy $\text{Tr}\left[\hat{H}_{AB} \hat{\rho}^p_{AB} \right]$. Using the marginals {$\hat{\rho}_{A(B)} = \text{Tr}_{B(A)}\left[|\psi\rangle\langle\psi|_{AB}\right]$} and simplifying the algebra, the expression for local ergotropy becomes 
\begin{align}
\mathscr{E}^l_{\hat{H}_{AB}}({\ket{\psi}_{AB}})&:= \text{Tr}\left [\hat{H}_{AB} \hat{\rho}_{AB} \right]- \min_{\substack{\hat{U}_{A} \in \mathscr{H}_A \\ \hat{U}_{B} \in \mathscr{H}_B}} \left(\text{Tr}\left[ \hat{H}_{A} \left( \hat{U}_{A}\hat{\rho}_{A} \hat{U}_{A}^\dagger \right) \right]\right.    \nonumber\\ 
&\hspace{2.5cm}+\left.\text{Tr}\left[ \hat{H}_{B} \left( \hat{U}_{B}\hat{\rho}_{B} \hat{U}_{B}^\dagger \right) \right]\right. \nonumber\\
&\hspace{.2cm}+ \left.\text{Tr}\left[ \hat{H}^{int}_{AB} \left( \hat{U}_{A}\otimes\hat{U}_{B}\hat{\rho}_{AB} \hat{U}_{A}^\dagger \otimes\hat{U}_{B}^\dagger \right) \right]\right). \label{Eq:locErg}
\end{align}
Two simplifications can be made at this point without loss of generality. First, {since $\ket{\psi}_{AB}$ is pure}, the passive state is the ground state and its energy can be set to zero. As such, the second term in Eq.~\eqref{Erg_global} vanishes. Secondly, the interactions can be quenched or only local measurements performed, which ensures the last term in Eq.~\eqref{Eq:locErg} is also absent. Therefore, the ergotropic gap reduces to 
\begin{equation}
\Delta_{\hat{H}_{AB}}^{A|B}({\ket{\psi}_{AB}}) = \text{Tr}\left[ \hat{H}_{A} \hat{\rho}^p_{A} \right]+\text{Tr}\left[ \hat{H}_B \hat{\rho}^p_{B} \right],
\label{Ergo-gap}
\end{equation}
where $\hat{\rho}^p_{A}$ and $\hat{\rho}^p_{B}$ are the passive states in the respective bipartitions. While the first simplification is intuitive, the second is particularly revealing: entanglement is fundamentally a property of the quantum state {$\ket{\psi}_{AB}$}, captured by the difference between global and local ergotropies, and does not explicitly depend on the structure of the Hamiltonian. Consequently, even a local Hamiltonian can faithfully capture the ergotropic gap and quantify entanglement. 
{In Appendix~\ref{App:newEg}, using small interacting spin models, we show how our  reformulation is consistent with the original measure in Eq.~\eqref{Eq:ErgGap} in regimes where it is valid, and also highlight how the new approach succeeds in regions where the original measure fails. 
Moreover, in Appendix~\ref{App:GGM}, the new operational thermodynamic measure is compared and shown to be consistent with established computational measures of genuine multipartite entanglement, such as the generalized geometric measure~\cite{SenDe2010} and multipartite concurrence~\cite{Ma2011}.}

For a pure biseparable state the marginals $\hat{\rho}_{A}$ and $\hat{\rho}_{B}$ in Eq.~\eqref{Ergo-gap} are also pure and the passive state energy is equal to the local ground state energy. Furthermore, for quenched systems or local measurements, the ground state is always product, and we set the ground state energy to zero. The ergotropic gap for such states thus always become zero. Whereas for entangled states the gap become finite, thereby certifying the presence of entanglement in the state.

Suitably defined functions of ergotropic gap can be obtained to certify genuine entanglement in multipartite states. One such quantity is the ergotropic volume \cite{Puliyil2022}. An $N$-partite quantum systems can be partitioned into two non-zero and disjoint groups of subsystems in total $D=2^{(N-1)}-1$ different ways. We denote  the partition as $\mathcal{P}~\&~\mathcal{P}^c$ s.t. $\mathcal{P}\cap\mathcal{P}^c=\emptyset~\&~\mathcal{P}\cup\mathcal{P}^c=\{1,\cdots,n\}$. For a $n$-partite pure state $\ket{\psi}$ the ergotropic volume is defined as the volume of an $D$-edged hyper-cuboid with sides $\Delta^{\mathcal{P}|\mathcal{P}^c}_{\hat{H}_0}(\psi)$, i.e
\begin{equation}
\Delta^V_{\hat{H}_0}(\psi) = \left(\prod_{\mathcal{P}=1}^D \Delta^{\mathcal{P}|\mathcal{P}^c}_{\hat{H}_0}(\psi) \right)^{1/D},
    \label{Eq:ErgVolDef}
\end{equation}
where $\hat{H}_0=\hat{H}_{\mathcal{P}}+\hat{H}_{\mathcal{P}^c}$ is the quenched Hamiltonian.  Using convex roof extension, this measure can further be extended to certify GME of mixed states as well. However, in this study we will limit ourselves to only stationary and time-evolved pure states of systems with strong interactions.

\section{Entanglement in Strongly Interacting Systems\label{Sec3}}

\subsection{Tavis-Cummings Model}
The Tavis-Cummings (TC) model~\cite{Tavis1968}, is an ideal theoretical platform to study strong interaction of quantum light with matter. 
{The model describes an ensemble of atoms interacting with each other only via a cavity field, with the total number of excitations conserved at all times in the system (Fig.~\ref{fig:FullErg}(e)).} The TC Hamiltonian is given by
\begin{equation}
    \hat{H}_{\text{TC}}^{N} = \omega_c \hat{a}^\dagger \hat{a} + \omega_a \sum_{i=1}^N \hat{\sigma}^i_z + {g \over \sqrt{N}} \sum_{i=1}^N (\hat{a}\hat{\sigma}_{+}^i + \hat{a}^\dagger \hat{\sigma}_{-}^i),
    \label{Ham:TC}
\end{equation}
where $\hat{a}$ and $\hat{a}^\dagger$ are the photon annihilation and creation operators, {with commutator $[\hat{a},\hat{a}^\dagger]=1$. 
The cavity and atomic transition frequencies are $\omega_c$ and $\omega_a$, respectively, with $g$ denoting} the strong coupling between the atoms and cavity photons. {Here,}
$\hat{\sigma}_{\alpha}^{i}$ with $\alpha \in \{x,y,z\}$ are the Pauli spin operators acting on $i^{\text{th}}$ atom,
with $\hat{\sigma}_{\pm}^i= \hat{\sigma}_x^{i} \pm i \hat{\sigma}_y^i$.
The model exhibits symmetries that help simplify its analysis. Conservation of total number of excitations $N_{ex} = \hat{a}^\dagger \hat{a} +\sum_{i=1}^N \hat{\sigma}^i_z$ is associated with $U(1)$ gauge symmetry. Spontaneous breaking of this symmetry leads to a normal to a superradiant phase transition, which occurs at the critical coupling of $g=\sqrt{\omega_c \omega_a}$ \cite{Kirton2019}. Notably, the ground state of the system is entangled in the superradiant phase. 

For an ensemble of identical spins, the Hamiltonian commutes with collective spin operators $\hat{J}^2$ and $\hat{J}_z$, where $\hat{J}_z = \sum_{i=1}^N \hat{\sigma}^i_z$. In the eigenbasis of the collective operators $\{\ket{J,M}\}$, the Hamiltonian is block-diagonal. As a result, in absence of any decoherence process, the unitary dynamics does not change total angular momentum of the state and effective dimension of the system is reduced from $2^N$ to $2J+1$ where $J$ is the total angular momentum of initial state. {Here, we focus on the $J=N/2$ subspace, spanned by the Dicke states $\{\ket{N/2,M},~\forall ~M\in -N/2 \cdots N/2\}$~\cite{Stockton2003,Chase2008}.} 
\begin{figure}
    \centering
    \includegraphics[width=\linewidth]{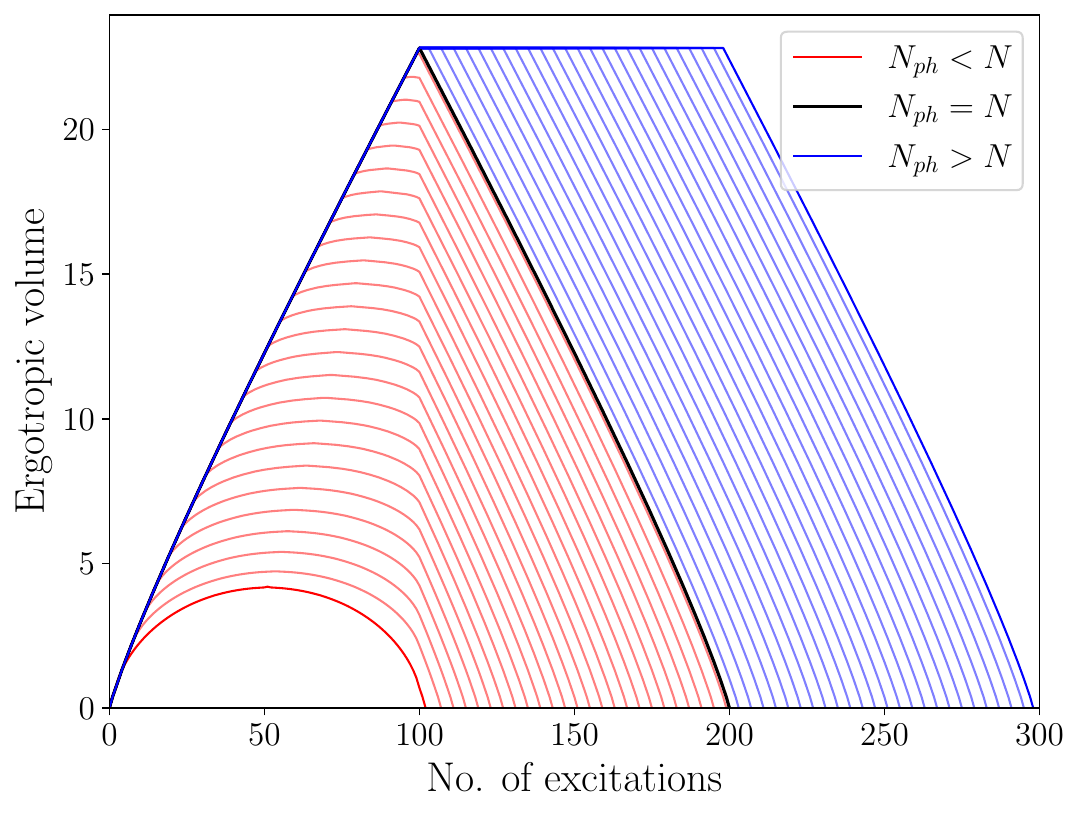}
    \caption{Ergotropic volume for dressed states with different number of excitations $i$. The system contains $N=100$ spins and maximum number of photons in the cavity varies from $N_{ph}=2$ to $198$, with $N_{ph}<N$ shown in red and $N_{ph}>N$ shown in blue.}
    \label{fig:TC}
\end{figure}

{The goal now is to compute {genuine multipartite} entanglement in a hybrid or dressed state of $N$ atoms and the cavity with total $i$ excitations $\hat{\rho}^i_{c,N}=|\Psi\rangle\langle\Psi|^i_{c,N}$}. These states are of particular importance as they are the eigenstates of the system Hamiltonian~\cite{Chiorescu2010}. For simplicity we consider an equal superposition of all the dressed states with $i$ excitation, i.e.,
\begin{equation}
    \ket{\Psi}_{c,N}^i = {1\over \sqrt{\mathcal{N}_i}} \sum_{l=0}^{N} \Theta_{N_{ph}}(i,{l}) \ket{i-{l}}\ket{{N\over 2},{l}-{N\over 2}},
    \label{Eq:DressedSt}
\end{equation}
where $\Theta_{N_{ph}}(i,l) = 1$ iff $i-l \leq N_{ph}$ otherwise it is 0 and $\max{(i)} = N+N_{ph} $. Here $N_{ph}$ is the maximum number of allowed photons in the cavity and $\mathcal{N}_i$ ensures proper normalization of the state with $i$ excitations.

Now to compute ergotropy, we need the eigenvalues of the marginals of this state across all bipartitions. Since across any bipartition, Schmidt coefficients of both the marginals are same, we focus only on the marginal which contains just the spin part of the system. Then, in the Dicke subspace we can write the marginal as $\hat{\rho}_n^i = \text{Tr}_{N-n}[\text{Tr}_c[\hat{\rho}_{c,N}^i]]$, which has matrix elements given by~\cite{Stockton2003}
\begin{equation}
    \hat{\rho}_n^i = {1\over \mathcal{N}_i}\sum_{l=0}^{n}\sum_{j=0}^{i-l} {C_{N-n,j}~C_{n,l} \over C_{N,l+j}} \ket{{n \over2},l-{n \over2}}\bra{{n \over2},l-{n \over2}},
    \label{Eq:RedDickeSt}
\end{equation}
where $C_{n,r} = n!/(r! (n-r)!)$ and $l = 0,1,\cdots n$. Computing marginals for $n=1,\cdots,N$ in this way gives reduced states for all possible bipartitions in the system, which is needed for computing ergotropic volume. The marginals which we obtain now, have become diagonal, making the calculation of eigenvalues and hence ergotropic volume easier.
Using these equations, we can now calculate the ergotropic gap, {using only the local or non-interacting terms of the Hamiltonian in Eq.~\eqref{Ham:TC}},
\begin{equation}
    \Delta_{\hat{H}^N_{\text{TC}}}^{n|c,N-n}(\hat{\rho}_{c,N}^i) = \text{Tr}\left[ \hat{H}_n (\hat{\rho}^i_{n})^p \right]+\text{Tr}\left[ \hat{H}_{c,N-n} (\hat{\rho}^i_{c,N-n})^p \right],
\end{equation}
where $\hat{H}_n = \omega_a \sum_{i=1}^n \hat{\sigma}^i_z$ and $\hat{H}_{c,n} = \omega_c \hat{a}^\dagger \hat{a} + \omega_a \sum_{i=1}^n \hat{\sigma}^i_z$. Finally, using Eq.~\eqref{Eq:ErgVolDef} we can write ergotropic volume of the state as,
\begin{equation}
    \Delta^V_{\hat{\rho}_{c,N}^i} = \left(\prod_{n=1}^{N} \Delta_{\hat{\rho}_{c,N}^i}^{n|c,N-n} \right)^{1/N},
\end{equation}
Figure~\ref{fig:TC} shows how GME varies as the number of excitations is increased in the dressed state of Eq.~\eqref{Eq:DressedSt}. From Eq.~\eqref{Eq:RedDickeSt}, it is evident that increasing the excitation number $i$ generally increases the number of levels populated in the marginal $\hat{\rho}_n^i$, as a result, ergotropic volume increases. However, this strictly happens only till $i <= \min{(N,N_{ph})}$, beyond this point if $N_{ph}<N$, ergotropy further increases till it reaches maximum at about $i\approx (N+N_{ph})/2$ as there are still higher excitation Dicke state or energy levels available in the system. But for $N_{ph}>=N$, there are no more levels remaining and the reduced state after tracing out the cavity is maximally entangled and unchanged till $i=N_{ph}$. Beyond this the GME starts reducing as the number of populated Dicke states starts reducing, right up to where the excitation number is  highest i.e., $i=N+N_{ph}$. At this point
only the completely separable or product state remains.

\subsection{Three-level Dicke model}
In the past decade, three-level Dicke model has gained considerable attention both from theoretical and experimental perspective. This is because these systems posses rich phase diagrams~\cite{Hayn2011,Zhang2017}, exhibit chaos~\cite{Stitely2020}, and have potential applications in microscopy, information storage and lasing~\cite{Lin2022,Fan2023}. Since, these systems host a relatively complex phase diagram, they are one of the best models to test the efficacy of ergotropic volume in differentiating the different {entangled} phases. As such, we consider an ensemble of $N$ identical, $V$-shaped three-level atoms coupled to a cavity of frequency $\omega_c$ (see Fig.~\ref{fig:FullErg}(f)). The atoms have two degenerate energy levels $\ket{1}$ and $\ket{2}$ that are separated from the ground state $\ket{0}$ by {transition} frequency $\omega_a$. The cavity couples the ground state and the two excited states $\ket{1}$ and $\ket{2}$ via two orthogonal cavity fields of strengths $g_1$ and $g_2$, respectively. The Hamiltonian of this system is given by~\cite{Lin2022},
\begin{eqnarray}
    \hat{H}_{\text{D}}^{N} &=& \omega_c \hat{a}^\dagger \hat{a} + \omega_a (\hat{\mathcal{A}_{11}} + \hat{\mathcal{A}_{22}}) + \frac{ig_1}{\sqrt{N}} (\hat{a} - \hat{a}^\dagger) \nonumber \\
    &\times& (\hat{\mathcal{A}_{01}} + \hat{\mathcal{A}_{10}}) + \frac{ig_2}{\sqrt{N}} (\hat{a} + \hat{a}^\dagger)(\hat{\mathcal{A}_{02}} - \hat{\mathcal{A}_{20}}),
    \label{Ham:Dicke}
\end{eqnarray}
where $\hat{\mathcal{A}}_{ij} = \sum_{k=1}^{N} |i_k\rangle\langle j_k|$ with $i,j =\{0,1,2\}$, and $\ket{j_k}$ denotes the $j$th level of the $k$th atom. {Again, $\hat{a}$ and $\hat{a}^\dag$ are the annihilation and creation operators for the cavity photons. The three-level Dicke model} possesses $\mathscr{Z}_2 \times \mathscr{Z}_2$ symmetry, which can be broken separately~\cite{Lin2022,Fan2023}. For $\max{(g_1,g_2)} < g_c \equiv \sqrt{\omega_c \omega_a}/2$, the system stays in normal phase ($NP$) with all atoms in ground state i.e., $\ket{\Psi_0} = \ket{0}_c\ket{00\cdots0}_N$. Above the critical coupling when $g_1 > g_2~ (g_2>g_1)$, the cavity is coherently populated and the parity symmetry is spontaneously broken, which leads to the occupation of $\ket{0}$ and $\ket{1}$ (or $\ket{2}$) levels and phase transition to superradiant phase $SP_1$ (or $SP_2$). In addition to this when $g_1=g_2 > g_c$, the $\mathscr{Z}_2 \times \mathscr{Z}_2$ symmetry gets enlarged to give rise to U(1) symmetry and all the three levels are populated in this case~\cite{Lin2022}.
\begin{figure}[t]
    \centering
    \includegraphics[width=\linewidth]{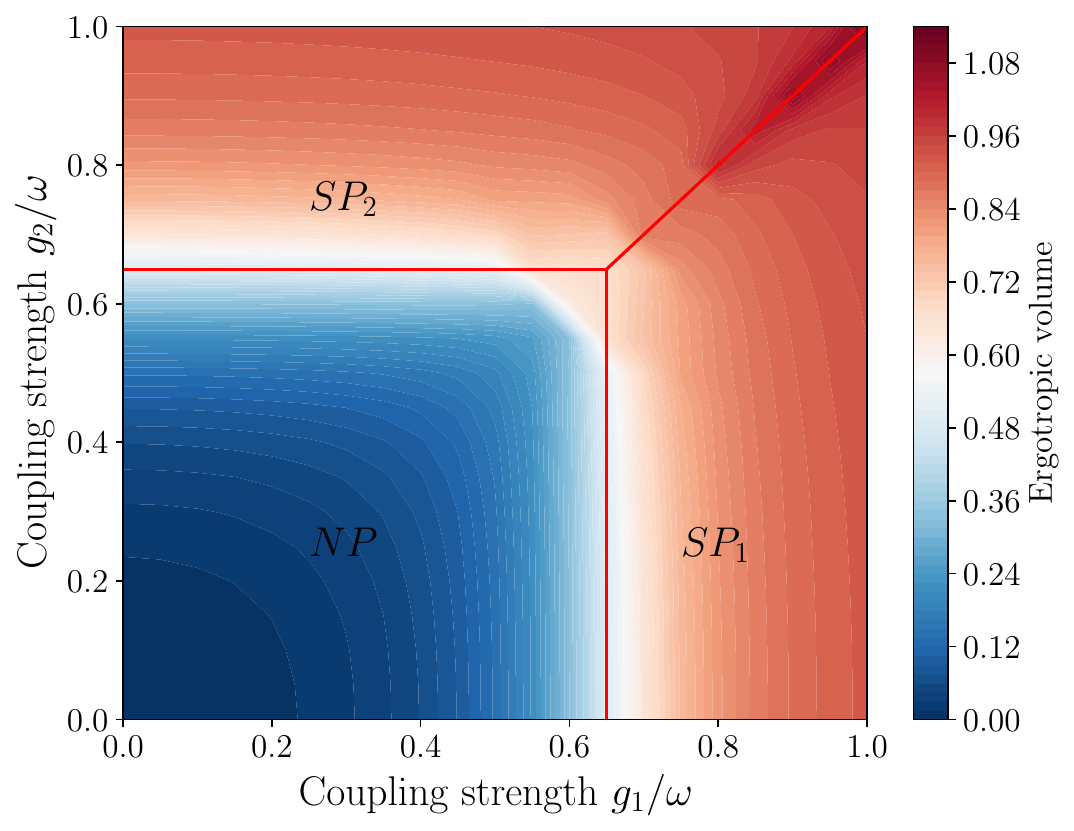}
    \caption{Phase diagram of a closed three-level system captured using ergotropic volume $\Delta^V_{\hat{H}_0^N}(\ket{\psi_N^{0}})$ of ground state $\psi_N^{0}$. The system parameters are $N=5$ and $\omega_c = \omega_a = \omega$.}
    \label{fig:Spin1Dicke}
\end{figure}

{Figure~\ref{fig:Spin1Dicke} shows the phase diagram of the system captured using ergotropic volume of the ground state.} Presence of different states in each phase gives rise to a rich entanglement landscape in this system. Normal phase basically contains only product state and therefore has zero entanglement, where as both superradiant phases have population in only two of the levels i.e., either $\{\ket{0},\ket{1}\}$ or $\{\ket{0},\ket{2}\}$, which gives rise to non-zero entanglement in these states. 
{For estimating the ergotropic gap and volume, only the non-interacting terms in Eq.~\eqref{Ham:Dicke} of the quenched Hamiltonian are used i.e., we set the interactions $g_1=g_2=0$. The gap is then numerically estimated across all bipartitions for a finite number of atoms.} 
As expected, both phase transitions from $NP$ to $SP_1$ and $SP_2$ are captured at critical coupling of $g_c \approx 0.65 \omega$. However, due to finite size effects, {it is away from the theoretically predicted value of $g_c = \omega/2$~\cite{Hepp1973}, which is achieved for large $N$ or at the thermodynamic limit.}
In addition to the two phase transitions, the $SP_1$ to $SP_2$ phase transition in regions away from ``triple point", i.e., where $g_1=g_2 > g_c$,
{is also captured by ergotropic volume. The line corresponds to fixed point bifurcations~\cite{Hines2005}, and exhibits a sharp increase in genuine multipartite entanglement. }

\subsection{Transverse field Ising model}
The transverse field Ising model (TFIM) has been a subject of interest for six decades since its inception in 1963 to model order-disorder transitions in potassium dihydrogen phosphate crystals~\cite{Gennes1963}. 
{It} is a simple model {that} describes an ensemble of spins, arranged in a lattice, {with interaction $J$ between any two nearest-neighbor spins.}
An external magnetic field of strength $h$ is applied {in the transverse direction} (Fig.~\ref{fig:FullErg}(d)).
{An} exciting feature of TFIM is that the one-dimensional (1D) model {exhibits a} quantum phase transition~\cite{Sachdev2011}. The Hamiltonian for a chain of $N$ spin$-1/2$ particles is given by
\begin{equation}
    \hat{H}_{\text{I}}^{N} = - \left( \sum_{i=1}^{N} \hat{\sigma}_z^i + g \sum_{i=1}^N  \hat{\sigma}_x^i\hat{\sigma}_x^{i+1} \right),
    \label{Eq:TFIM}
\end{equation}
where $g = J/h$ is the relative strength of interaction compared to the external field and $\hat{\sigma}_{\alpha}^{i}$ are {once again} the Pauli operators acting on $i^{\text{th}}$ spin. 

At the critical value of $g=1$, this model undergoes quantum phase transition from an ordered phase to a disordered phase \cite{Fradkin2013}. Ordered phase corresponds to the case where ground state $\hat{\rho}_N^0$ violates spin-flip symmetry (all spins aligned along +z or -z axis at $g=0$) in contrast to disordered phase where this symmetry is preserved (all spins aligned along +x axis at $g\rightarrow \infty$). This phase-transition point is also captured by the entanglement in the ground state~\cite{Osborne2002,Hofmann2014}, which makes it one of the ideal model to test ergotorpic volume as a measure of {multiparty} entanglement.

Using Jordan-Wigner transformation \cite{Jordan1928}, the Hamiltonian in Eq.~\eqref{Eq:TFIM} can be mapped to a system of non-interacting fermions, which in momentum space reads \cite{Bhat2024,mbeng2020},
\begin{equation}
    \hat{\mathbb{H}}_{\nu=1} = 2 \sum_{k} (1-g)\cos{k}~\hat{\sigma}_z - g \sin{k}~ \hat{\sigma}_y.
    \label{Eq:kspaceTFIM}
\end{equation}
For this Hamiltonian, we can compute the eigenvalues ($\lambda_n^p$) of marginals ($\hat{\rho}^0_M = \text{Tr}_{N-M}[\hat{\rho}_N^0],M\leq N $) of the ground state ($\hat{\rho}^0_N$) using the eigenvalues of one-point correlation matrix as~\cite{Sirker2014} (see Appendix~\ref{App:MarGS_TFIM} for more details),
\begin{equation}
    \lambda_n^p = {1 \over \mathcal{Z}} \prod_q \left({1\over \zeta_q^n}-1\right)^{f_q^{(p)}},
    \label{lambda}
\end{equation}
where $f_q^{(p)} \in \{0,1\}$ are the fermion occupation number at each site $q \in 1, 2,\cdots n$ and $\mathcal{Z}$ is normalisation such that $\text{Tr}[\hat{\rho}_n^0]=1$. Using these eigenvalues, 
{the ergotropic gap can be readily calculated to give
\begin{equation}
    \Delta^M_{\hat{H}_0^M}(\hat{\rho}_N^0) = \text{Tr}\left[ H_{0}^M \hat{\rho}^p_{M} \right]+\text{Tr}\left[ H_{0}^{N-M} \hat{\rho}^p_{N-M} \right],
    \label{delta_m}
\end{equation}
where {$\hat{\rho}^p_{M}$ and $\hat{\rho}^p_{N-M}$ are the passive states corresponding to the complementary partitions.
The genuine multipartite entanglement as captured by the} ergotropic volume is then given by,
\begin{equation}
    \Delta^V_{\hat{H}_0^N}(\hat{\rho}_N^0) = \left(\prod_{M=1}^{N/2}  \Delta^M_{\hat{H}_0^M}(\hat{\rho}_N^0) \right)^{2/N}.
    \label{delta_v}
\end{equation}
\begin{figure}
    \centering
    \includegraphics[width = \columnwidth]{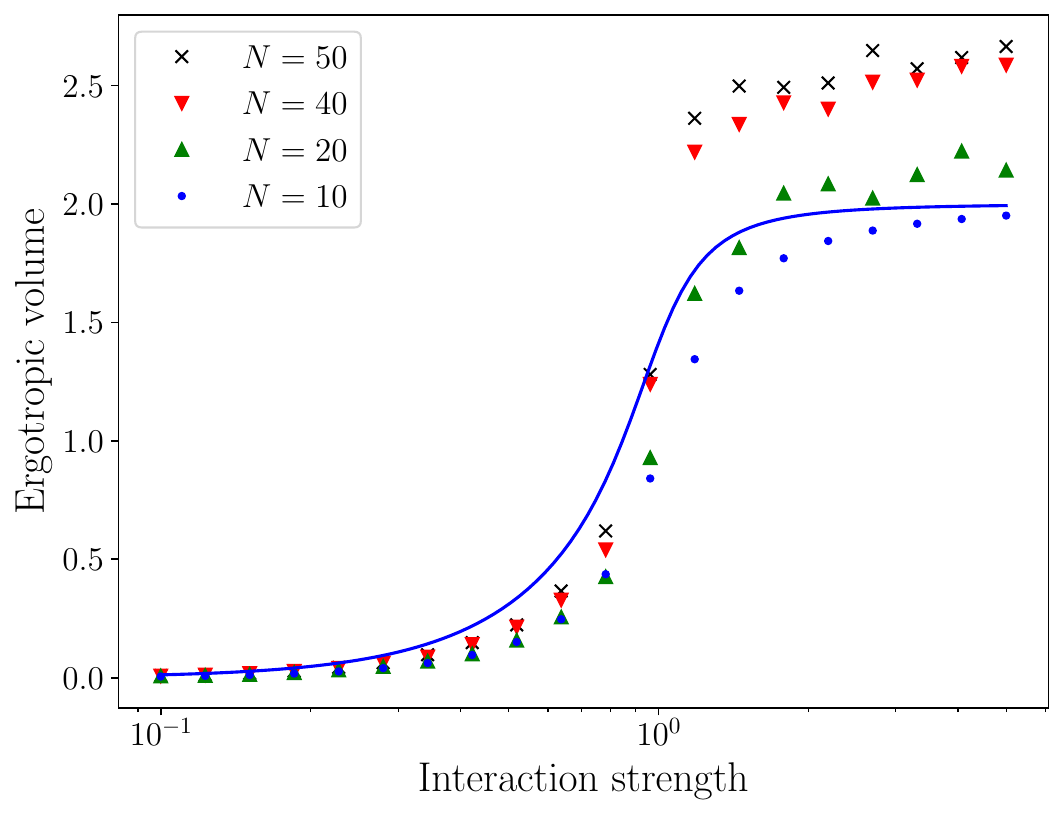}
    \caption{Ergotropic volume for ground state of transverse field Ising model. The markers represent the ergotropy computed using the analytical method for $N=10$ (blue dots), $20$ (green up-triangles), $40$ (red down-triangles) and $50$ (black cross) spins. For $N=10$ spins, the solid blue curve represents the ergotropic volume computed numerically considering all possible bipartitions.}
    \label{fig:AnaIs}
\end{figure}

{Figure~\ref{fig:AnaIs} shows the genuine multipartite entanglement of the ground state of the transverse Ising model, as captured by the ergotropic volume,  as the interaction term $g$ is varied.
The ergotropic volume captures the quantum phase transition, which is determined by the inflection point i.e. the point where double derivative of entanglement with respect to $g$ is 0. As predicted by theory~\cite{Suzuki2013}, this happens at $g_c = 1$ for different values of $N$ and ergotropic volume captures this point very well.} 

Note that analytical expressions for calculating the eigenvalues of the marginals in Eq.~\eqref{lambda}, are only for partitions across contiguous blocks. Therefore, non-contiguous partitions do not contribute to the calculation of ergotropic gap and volume in Eqs.~\eqref{delta_m}-\eqref{delta_v}.
However, biseparability across non-contiguous blocks is fairly low due to the symmetry of the Hamiltonian i.e., the states of the Hamiltonian is more likely to factorize into contiguous blocks rather than random partitions. As such, the ergotropic volume computed using analytical calculations will only be a lower bound.
{This is supported by numerical calculation of ergotropic volume, as shown in Fig.~\ref{fig:AnaIs} for $N=10$ (solid blue line), where all bipartitions have been considered. These numerical values upper bound the analytically obtained ergotropic volume (blue dots). As continguous bipartitions increase with higher $N$, this bound is expected to get tighter. However, this does not qualitatively affect our overall results.}


\section{Implementation in Quantum circuits}\label{Sec4}
\begin{figure*}
    \centering
    \includegraphics[width=\linewidth]{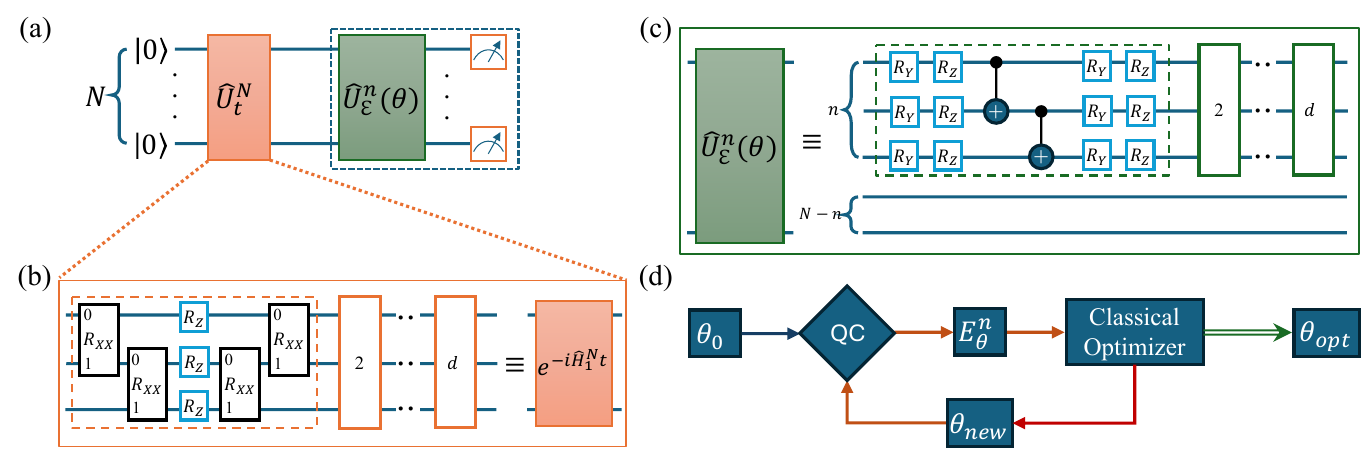}
    \caption{Measuring ergotropy using quantum circuits. (a) The quantum circuit for studying the entanglement dynamics in the system. (b) Unitary $\hat{U}_t$ time evolves the initial state (ground state in this case), which is implemented using Suzuki-Trotter decomposition~\cite{Berry2007}, where the base circuit (orange dashed box) is repeated $d$ times to reduce error. (c) Ansatz circuit for finding the optimal unitary $\hat{U}_{\mathcal{E}}^n(\theta_{opt})$ that minimizes the energy of the time-evolved state. This circuit also has a depth $d$ to reduce the error during computation. (d) Flow-chart for passive state optimization protocol, where quantum circuit computes the energy of state $E_{\theta}^n$ after implementing unitary $\hat{U}_{\mathcal{E}}^n(\theta)$. This energy is then fed to a classical optimizer which keeps on changing $\theta$ and running the quantum circuit in (a) until the optimal value $\theta_{opt}$ that minimizes the energy is found.}
    \label{fig:QC}
\end{figure*}

{The key strength of ergotropic volume as an entanglement measure is that it can be readily measured in experiments where energy observables are more accessible, including quantum circuits implemented in noisy intermediate scale quantum (NISQ) devices~\cite{Preskill2018,Bharti2022}.} 
In fact, ergotropy of quantum states were recently measured in NMR experiments using optimal control~\cite{Joshi2024a,Joshi2024b} and in quantum circuits using variational quantum algorithms~\cite{Hoang2024}.
{This makes ergotropic volume a powerful tool in physical implementation of} quantum information and communication, as several protocols depend on entanglement as a key resource~\cite{Horodecki2009, Chitambar2019}. 
{However, the role of multipartite entanglement in quantum computation such as measurement based computation~\cite{Raussendorf2001} and distributed quantum computing~\cite{Grover1997, Cirac1999,Jiang2007} has always been more intriguing. As such, the question of whether the state during any stage of quantum computation or quantum simulation is multiparty entangled or not can be of fundamental significance.} 

The ergotropic volume of a quantum state can be estimated using {a quantum circuit by measuring} the bipartite ergotropic gap, as defined in Eq.~\eqref{Eq:ErgGap}. {The approach is based on variational measurement of global and local observables in the circuit, without relying on partial or full quantum state tomography.}
{Importantly, measurement across a bipartition $A:B$ requires two minimizations to compute the passive state energy corresponding to the marginals or reduced state of the subsystems $A$ and $B$.} Variational quantum algorithms can be used for these minimization tasks on present NISQ architecture. 
{As illustrated in Fig.~\ref{fig:QC}, the protocol} makes use of a standard classical-quantum feedback loop~\cite{Cerezo2021} (see Fig.~\ref{fig:QC}(d)) to find the {optimal} parameters of a parametrized quantum circuit Fig.~\ref{fig:QC}(c), which {then} minimizes the energy of the input quantum state Fig.~\ref{fig:QC}(a-b). 
The quantum part of the algorithm evaluates the quantum circuit to compute the energy of the output state and the classical optimizer is used as a feedback to keep on changing the parameters till the optimal parameters are found. 

As an example, consider a system of $N$ qubits which are interacting via {TFIM Hamiltonian} in Eq.~\eqref{Eq:TFIM}. The time evolved states of {this many-body interacting system} are typically multiparty entangled and the circuit can serve as state preparation for some quantum information protocol. 
Therefore, we want to study the entanglement dynamics of the system {and show that arbitrary entanglement in the state can be estimated during any phase of the computation.}
The first step in {the quantum simulation is to reach an arbitrary} time evolved state of the system. 
{Starting} with all qubits in state $\ket{0}$, the system evolves under the unitary $\hat{U}_t = \text{Exp}\left[- i \hat{H}_1^N t  \right]$, as follows,
\begin{equation}
    \ket{\psi(t)} = \hat{U}_t \ket{\psi(0)} = e^{- i \hat{H}_1^N t} \ket{0}^{\otimes N}
\end{equation}
Ideally, this can achieved {on a gate-based quantum circuit} using Suzuki-Trotter decomposition~\cite{Hatano2005}, which decomposes the global time evolution unitary into local gates as shown in Fig.~\ref{fig:QC}(b). 

However, the number of gates required grows with both time $t$ and number of qubits $N$. This is not suitable for noisy hardware as beyond a certain time and number of qubits
{the simulation is not accurate anymore.}
To overcome this, we use the approximate quantum compilation algorithm AQCtensor~\cite{Robertson2025}, to efficiently approximate the time evolved state {by making} use of tensor network methods to speed up the optimization. 
In this method, time evolution is simulated with much shorter circuits than standard Trotterization. Instead of running a deep Trotterized circuit for the whole time, it is simulated classically using matrix product state representation, and a shallow parametrized circuit is trained to reproduce the time evolved state by maximizing fidelity between the two. This compressed circuit effectively “compiles” the long-time evolution into fewer layers i.e, it requires a circuit depth  {that is significantly lower} than the {first order} Suzuki-Trotter circuit {used in a typical variational quantum algorithm}. 

\begin{figure}
    \centering
    \includegraphics[width=0.925\columnwidth]{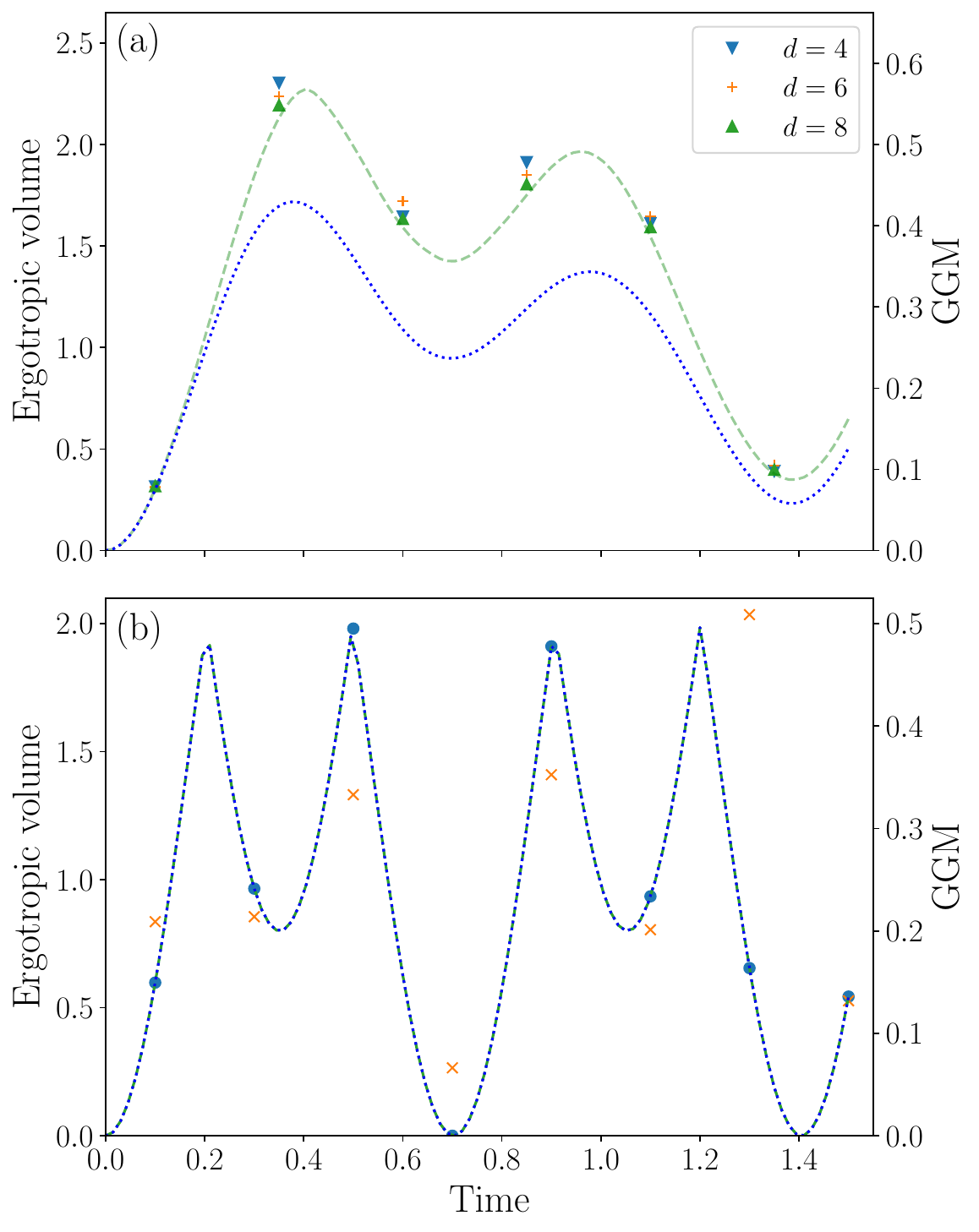}
    \caption{Measurement of ergotropic volume of time evolved state of TFIM using variational quantum algorithm. The system is a 1D spin chain with nearest neighbor interaction strength $g=2 h$ and (a) $N=6$ and (b) $N = 2$ spins. (a) Ideal simulations for $N=6$ spins for different circuit depth $d$ of the ansatz circuit in Fig.~\ref{fig:QC}(c).  (b) Comparison of the implementation on an ideal state vector simulator (blue dots) {with $d=2$ and same circuit implemented on a} noisy IBM fake-Perth simulator (orange cross), which is the noise model for real 7 qubit IBM-Perth quantum computer. In both simulations, the exact numerical results for ergotropic volume are shown as green-dashed curves and blue-dotted curves shows the corresponding values for generalized geometric measure of entanglement (GGM) on right y-axes.}
    \label{fig:QCsim}
\end{figure}

{The subsequent} step is to find out the unitary that minimizes the energy corresponding to the non-interacting Hamiltonian $\hat{H}_0^n$ of a bipartition $A:B$, where the party $A$ contains $n$ spins. This unitary $\hat{U}^n_{\mathcal{E}}(\theta)$ is a parametrized circuit composed of single-qubit ($Y$ and $Z$) and two-qubit ($CX$) gates, with a set of parameters $\theta$. However, these gates are only applied to the qubits which are present in the partition in consideration (see Fig.~\ref{fig:QC}(c)).
The energy of state after the unitary is then given by,
\begin{equation}
    E_\theta ^n = \bra{\psi(t)} \hat{U}^{n \dagger}_{\mathcal{E}}(\theta) \hat{H}_0^n  \hat{U}^{n}_{\mathcal{E}} (\theta) \ket{\psi(t)}.
    \label{Eq:ParEner}
\end{equation}
The goal here is to find a set of optimal parameters $\theta_{opt}$, which minimizes the energy above. On the quantum circuit, the energy in Eq.~\eqref{Eq:ParEner} can be computed by measuring the qubits after the unitary. This energy is the feedback to a classical optimizer~\cite{Nocedal2006}, which keeps on varying the parameters in the quantum circuit until the optimal parameters $\theta_{\text{opt}}$ are found. A flowchart describing these steps is shown in Fig.~\ref{fig:QC}(d). The  ergotropic gap across the bipartition $A:B$ can then be written as
\begin{equation}
    \Delta^{A|B}_{\hat{H_1^N}}(\ket{\psi(t)}) = E^n_{\theta_{opt}} + E^{N-n}_{\theta_{opt}}.
\end{equation}
The standard procedure to compute ergotropic volume would then involve computing geometric mean of  ergotropic gap across all possible bipartitions. However, considering all possible bipartitions is not explicitly required in the transverse field Ising model case, as we know that we can get a lower bound to ergotropic volume by just considering the contiguous blocks (refer to Fig.~\ref{fig:AnaIs}). As a result, we only need to measure the passive state energies for $n=1,\cdots,N-1$, which allows us to calculate ergotropic gap across all contiguous blocks. 

Though the entire procedure is relatively straight forward, it takes considerable effort to find the optimal parameters for which this method works. One such parameter is finding the optimal circuit depth, i.e., the number of repetitions $d$ of the base (parametrized ansatz) circuit in Fig.~\ref{fig:QC}(c). 
Figure~\ref{fig:QCsim}(a) shows the simulation of the entire algorithm described above on quantum circuit~\cite{Abhari2024} for $N = 6$ spin TFIM for different circuit depths. The ideal quantum circuit results eventually match the numerical calculations involving exact diagonalization. However, we find that in general we need more circuit depth to get better results as $N$ is increased. This is because increasing the circuit depth reduces the errors generated from variational algorithm used to compute passive state energy~\cite{Hoang2024}. We also compare the entanglement dynamics obtained from ergotropic volume with the one obtained from generalized geometric measure of entanglement (GGM) (see Appendix~\ref{App:GGM} for the definition). Both measures qualitatively suggests the same dynamics, but computing GGM generally requires full state tomography, which is not advantageous experimentally. On introducing noise into the system, we see deviation in the exact and simulated results (see Fig.~\ref{fig:QCsim}(b)). This is because of presence of both gate errors and errors inherent to the variational algorithm itself.

\section{Conclusion}\label{Sec5}

{In recent years, quantum thermodynamic quantities such as ergotropy have been used to define measures of entanglement that are experimentally more accessible and also suitable for implementation in quantum simulators and gate-based NISQ devices. In this work, we show how the formalism can be extended to measure multipartite entanglement in strongly interacting systems. The key strength of our approach is that the ergotropic measurements can be limited to quenched Hamitonian or local observables, which can be readily implemented in several experiments or quantum simulators.} 

{Our method allows us to derive analytical expressions for the ergotropic volume, which captures the genuine multiparty entanglement in stationary states of hybrid cavity QED systems and many-body spin models. These are consistent with numerical values obtained for other geometric measures of entanglement. 
Moreover, our measures consistently capture the physical properties of the state such as quantum phase transitions and critical behavior, highlighting its strength as a good figure of merit to study entanglement in strongly interacting systems. We also present a protocol demonstrating how the measurement of ergotropic gap and ergotropic volume can be integrated in a quantum circuit and estimated using variational quantum algorithms. This readily allows one to measure multipartite entanglement at different stages of a quantum simulation, especially while performing such simulations on NISQ devices.}

\section*{Acknowledgements}
H.S. acknowledges financial support from the Prime Minister’s Research Fellowship (ID: 1302055), Govt. of India. M. B. acknowledges funding support from the National Quantum Mission (NQM), an initiative of the Department of Science and Technology, Govt. of India. H.S.D. acknowledges support from SERB-DST, India under a Core-Research Grant (No: CRG/2021/008918) and from IRCC, IIT Bombay (No: RD/0521-IRCCSH0-001). This work was carried out under the collaborative framework of the Memorandum of Understanding between IIT Bombay and SNBNCBS, Kolkata. We acknowledge National Supercomputing Mission (NSM) for providing computing resources of `PARAM RUDRA' at S.N. Bose National Centre for Basic Sciences, which is implemented by C-DAC and supported by the Ministry of Electronics and Information Technology (MeitY) and Department of Science and Technology (DST), Government of India.

\appendix
\section{Ergotropic gap in interacting and quenched systems\label{App:newEg}}

{In this section, we explain how reformulating the definition of ergotropic gap using quenching or local measurement  is consistent with the genuine multipartite entanglement measure defined in Ref.~\cite{Puliyil2022}. We also highlight how our measure correctly estimates the entanglement when the original measure is not valid in interacting regimes with entangled ground states.}

{Let us consider simple two-spin models, where the optimisation of both local and global unitaries can be implemented with fairly high accuracy. The first is the Jaynes Cummings (JC) Hamiltonian~\cite{Jaynes1963}, with a single photon interacting with a two-level emitter, given by Eq.~\eqref{Ham:TC} for $N=1$ and $\omega_c=\omega_a = 1$. The second is the two-spin Ising model, given by Eq.~\eqref{Eq:TFIM} for $N=2$. Note that both models represent essentially two-qubits, with interaction parameter $g$, but exhibit two interesting regimes. For the JC model, the ground state (GS) is separable for $g<1$, and the original measure is valid. On the other hand, for Ising model the GS is always entangled for finite $g$ and as such the original measure is not valid. }  
\begin{figure}[]
    \centering
    \includegraphics[width=.9\columnwidth]{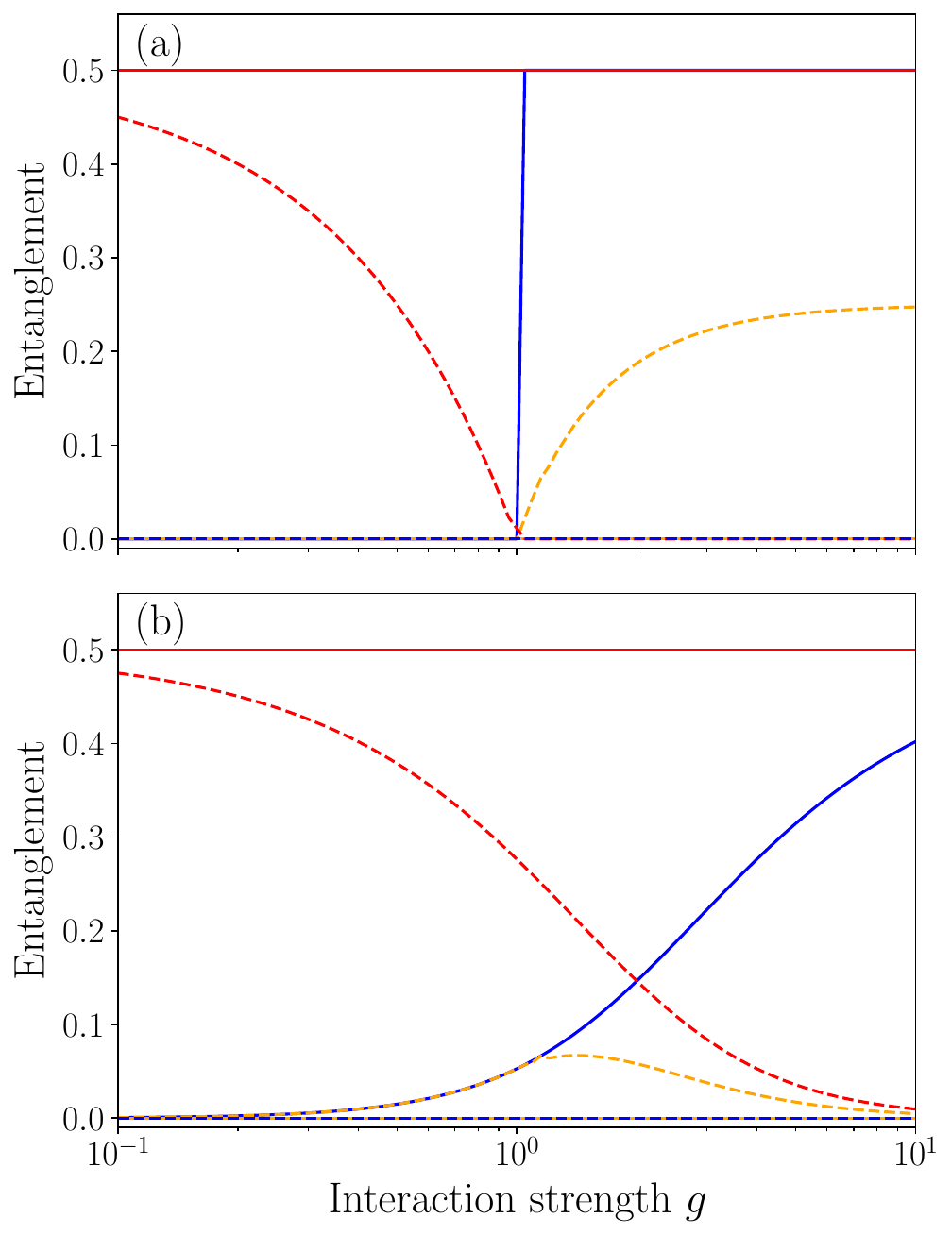}
    \caption{Ergotropic volume in interacting systems. The figure corresponds to a) a single photon and single emittter Jaynes-Cummings model b) two-spin transverse field Ising model. The plots show the entanglement measured for a product state $\ket{01}$ (yellow), a maximally entangled singlet state $(\ket{01}-\ket{10})/\sqrt{2}$ (red) and ground state (blue) of the system. The dashed curves represent the original definition of entanglement based on ergotropic volume, while the solid lines represent the reformulated approach based on quenched interactions.}
    \label{fig7}
\end{figure}

{Figure~\ref{fig7} summarizes results for three quantum states: a separable state (yellow), a maximally entangled singlet state (red), and the GS of the interacting Hamiltonians (blue). Since there is no closed-form expression for ergotropy for general interacting Hamiltonians, we numerically optimize over global and local unitaries to obtain the ergotropic volume. In special cases, for states with maximally mixed marginals such as the Bell or Werner states, the optimization can be performed analytically by computing parallel ergotropy~\cite{Castellano2025}.
The original definition is shown using dashed lines, with the reformulated measure given by solid lines.}
{For both the JC (Figs.~\ref{fig7}a) and Ising model (Figs.~\ref{fig7}b), the measures are consistent when the GS is separable i.e., $g < 1$ or $g \approx 0$, respectively. However, we can see that the original measure fails to correctly estimate entanglement when the GS is not separable. For instance, finite entanglement is detected even for the separable state (dashed-yellow) and the singlet (dashed-red) decreases to zero. Moreover, the GS entanglement (dashed-blue) is always zero and is never detected. However, the reformulated ergotropic volume, correctly estimates the entanglement in the system everywhere, for all the three states, including the ground state of the interacting Hamiltonians.}

\section{Other measure of genuine multipartite entanglement\label{App:GGM}}
Let us consider two archetypal examples of genuine multipartite entanglement (GME) measures i) Generalised geometric measure (GGM)~\cite{SenDe2010}, which is a computable geometric measure of GME and ii) an entropic measure called multiparty concurrence~\cite{Ma2011}. This is then compared with our thermodynamic measure based on ergotropic gap as defined in Eq.~\eqref{Eq:ErgVolDef}.

GGM is defined as the minimum distance of a given state from the set of all states that are not genuinely multiparty entangled. Mathematically, for a $n$-partite pure quantum state $\ket{\psi}_{A_1,A_2, ... A_n} \in  \mathcal{H}_A = \otimes_{i=1}^{n} \mathcal{H}_{A_i}$, the generalised geometric measure is defined as~\cite{SenDe2010,SRoy2019}
\begin{equation}
\mathscr{E}^G_{\ket{\psi}}= 1 - \Lambda_{max}^2(\ket{\psi})
\end{equation}
here, $\Lambda_{max}(\ket{\psi}) = \max|\bra{\phi}\ket{\psi}|$,
 where maximization is performed across all pure states $\ket{\phi}$ that are at least biseparable. It can further be written as,
\begin{equation}
\mathscr{E}^G_{\ket{\psi}}= 1 - \max({\lambda^2_{\mathcal{A|B}}})
\end{equation}
where $\lambda^2_{\mathcal{A|B}}$ are the Schmidt coefficient in the $\mathcal{A|B}$ bipartition of the state $\ket{\psi}$, with $\mathcal{A}\cup\mathcal{B}=\{A_1,A_2,\ldots,A_N\}$ and $\mathcal{A}\cap\mathcal{B}= \varnothing$. 
GGM is a simple but versatile measure, which is easy to calculate in small systems, and has been  a powerful tool in the study of multiparty entanglement in complex many body states such as dimer networks~\cite{Dhar2011,Dhar2014} or quantum phase transitions~\cite{Biswas2014}. 

On the other hand, the GME-concurrence for the state $\ket{\psi}$ is defined as~\cite{Ma2011} 
\begin{equation}
\mathscr{E}^C_{\ket{\psi}} = \min_{\rho_{A}} \sqrt{2[1-\textrm{Tr}(\rho^2_{A})]},
\end{equation}
where the optimization is over all possible marginals or reduced states $\rho_{A}$ in the partition $\mathcal{A|B}$. The multiparty entanglement here is related to the purity of the reduced state $\textrm{Tr}(\rho^2_{A})$, which is equal to unity if the state is separable across the partition. 

\begin{figure}
    \centering
    \includegraphics[width=.9\columnwidth]{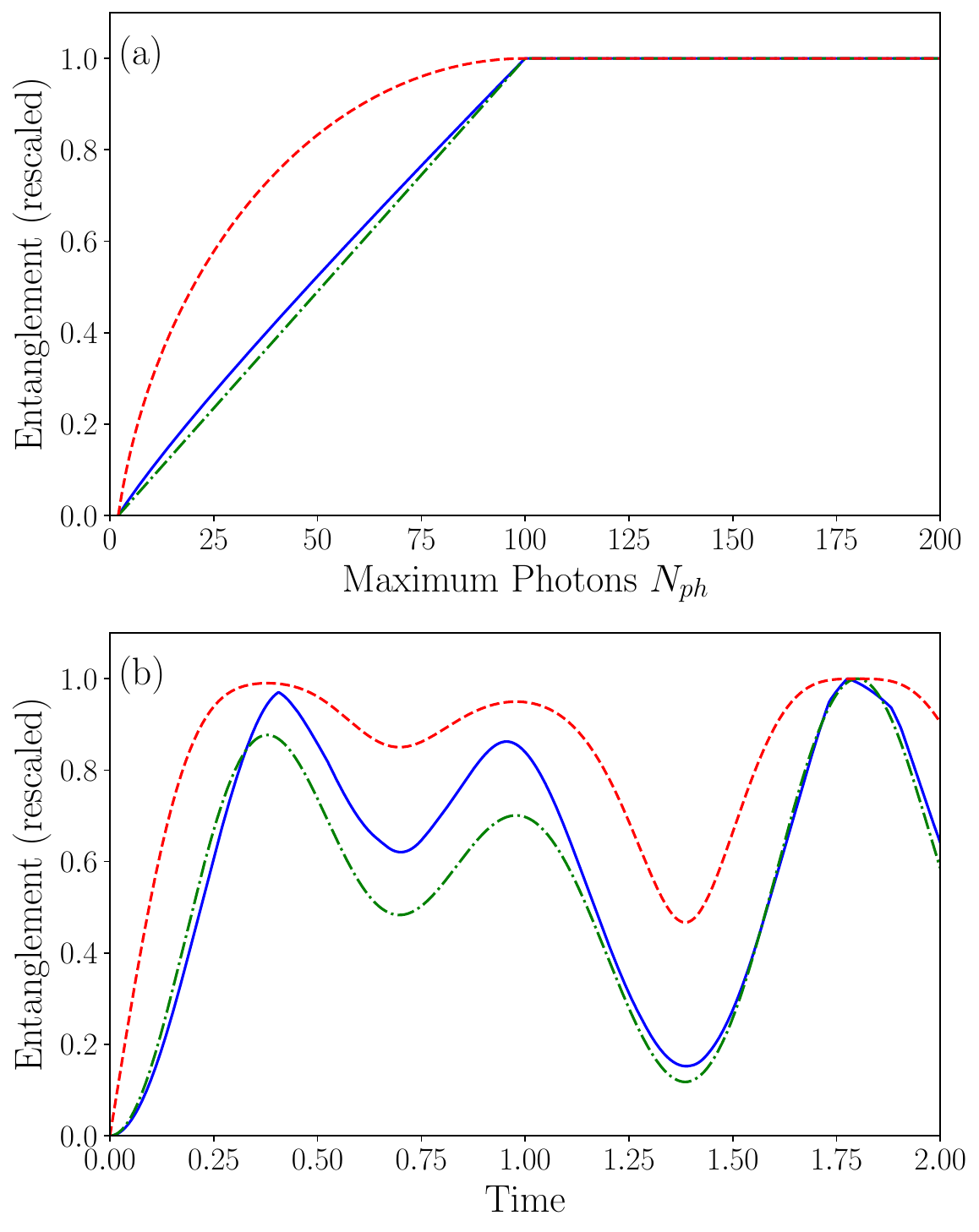}
    \caption{Benchmarking ergotropic volume with GGM and GME-concurrence. (a) Rescaled entanglement values of different measures of entanglement for dressed state in Eq.~\eqref{Eq:DressedSt} with $N = 100$ spins and $i = 100$ excitations, where maximum number of photons in cavity $N_{ph}$ is varied from $2$ to $200$. (b) Rescaled entanglement values for entanglement dynamics in transverse-field Ising model for $N = 6$ and $g = 2 h$. In both plots dotted (red), dot-dashed (green) and solid (blue) curve represents GME-concurrence, GGM and ergotropic volume, respectively.}
    \label{fig:Comp}
\end{figure}

{Figure~\ref{fig:Comp} shows the behavior of the three measures of GME for dressed states in Tavis-Cummings model (Fig.~\ref{fig:Comp}(a)) and for entanglement dynamics in transverse-field Ising model (Fig.~\ref{fig:Comp}(b)). To faithfully compare the three measures, the entanglement values in the three datasets were rescaled to lie between 0 and 1. Quantitatively the three measures show different behavior as mathematically GME-concurrence and GGM are based on extremal values i.e., minimum linear entropy or maximum fidelity with biseparable states, whereas
ergotropic volume is the geometric mean over all possible bipartitions. However, qualitatively all three entanglement measures show similar behavior as the system parameters are varied in complex many-body states.
}

\section{Marginals for ground state of TFIM}\label{App:MarGS_TFIM}
{The Jordan-Wigner transformation helps us to diagonalize the Hamiltonian into blocks of dimension 2 in momentum space. Using the eigen decomposition of these blocks, the ground state of the Hamiltonian in Eq.~\eqref{Eq:kspaceTFIM}, can be expressed as, 
\begin{equation}
\left|\psi_{0}\right\rangle=\prod_{k>0} a_{k} c_{k}^{\dagger} c_{-k}^{\dagger}|0\rangle+b_{k}|0\rangle.
\end{equation}
where $\hat{c}_{k}^\dagger$ is the fermionic creation operator in $k$-space and $a_k, b_k$ are the elements of the eigenvector corresponding to negative eigenvalue of $k$-momentum block.}

Now, we trace out the degrees of freedom that are not of interest, to obtain the marginals which can be expressed as an exponential of a free-fermion operator, also known as the entanglement Hamiltonian~\cite{Sirker2014}:

\begin{equation}
\rho_X = \frac{1}{\mathcal{Z}} \exp(-H_X) = \frac{1}{\mathcal{Z}} \exp \left( -\sum_{k=1}^{N} \epsilon_k c_k^\dagger c_k \right)
\end{equation}
The entanglement Hamiltonian $H_X$ is expressed as $H_X = -\sum_{k=1}^{N} \epsilon_k c^\dagger_k c_k$, where the $\epsilon_k$ are the eigenvalues of entanglement Hamiltonian.

Now, we go on to calculate the one-body correlation matrix, $\mathbf{C}_{\text{initial}}$, from the ground state.
The correlation matrix serves as an intermediate step in the calculation, instead of directly calculating the reduced density matrix, allowing us to obtain the entanglement spectrum and, hence, the eigenvalues of the reduced density matrix in an efficient way, which are required for computing the ergotropic gap. It has the following structure~\cite{Bhat2024}:

\begin{equation}
\mathbf{C}_{\text{initial}} = \begin{pmatrix}
\mathbb{C} & \mathbb{F} \\
\mathbb{F}^\dagger & \mathbb{I} - \mathbb{C}
\end{pmatrix}
\end{equation}
Specifically, the matrix elements are given by:
\begin{equation}
\begin{aligned}
\mathbb{C}_{ij} &= \langle c_i^\dagger c_j\rangle = \frac{2}{N} \sum_{k \in BZ/2} |a_k|^2 \cos(k(i-j)), \\
\mathbb{F}_{ij} &= \langle c_i^\dagger c_j^\dagger\rangle = \frac{2}{N} \sum_{k \in BZ/2} a_k^* b_k \sin(k(i-j)),
\end{aligned}
\end{equation}
where $i,j$ run over subsystem size $n<N$ and $\mathcal{K}$ represents half of the Brillouin zone. For a free fermion model, the eigenvalues $\lambda_n^p$ of marginals $\hat{\rho}_n^0 (= \text{Tr}_{N-n}[\hat{\rho}_N^0]$) can be computed easily using the eigenvalues $\zeta_q^n$ of correlation matrix using \cite{Sirker2014},
\begin{equation}
    \lambda_n^p = {1 \over \mathcal{Z}} \prod_q \left({1\over \zeta_q^n}-1\right)^{f_q^{(p)}}
\end{equation}
where $f_q^{(p)} \in \{0,1\}$ are the fermion occupation number at each site $q \in 1, 2,\cdots n$ and $\mathcal{Z}$ is normalisation such that $\text{Tr}[\hat{\rho}_n^0]=1$.


\bibliography{references}

\end{document}